\documentclass [12pt]{article}
\begin{document}

\thispagestyle{empty} \setcounter{page}{1}
\begin{titlepage}
\begin{flushright}
Imperial/TP/98--99/13.
\end{flushright}
\vskip 1cm
\begin{center}
{\large{\bf The action operator for continuous-time
histories }} \vskip 2cm K. Savvidou\footnote{e-mail:
k.savvidou@ic.ac.uk. } \vskip 0.4cm {\it Theoretical
Physics Group\\ Blackett Laboratory\\ Imperial College of
Science, Technology \& Medicine\\ London SW7 2BZ, U.K. }
\vskip 0.7cm  November 1998
\end{center}
\vskip 0.9cm
\begin{abstract}
We define the action operator for the consistent histories
formalism, as the quantum analogue of the classical action
functional, for the simple harmonic oscillator case. We
conclude that the action operator is the generator of time
transformations, and is associated with the two types of
time-evolution of the standard quantum theory: the
wave-packet reduction and the unitary time-evolution. We
construct the corresponding classical histories and
demonstrate the relevance with the quantum histories.
Finally, we show the relation of the action operator to the
decoherence functional.
\end{abstract}
\end{titlepage}
\renewcommand{\theequation}{\thesection.\arabic{equation}}
\let\ssection=\section
\renewcommand{\section}{\setcounter{equation}{0}\ssection}

\section{Introduction}
One of the basic elements in the consistent histories
formalism is the idea of a `homogeneous history'. This is a
time-ordered sequence of propositions about the system and,
in the original approaches to the formalism, is represented
by a class operator $\tilde{C}$
\begin{equation}
\tilde{C}:=
U(t_0,t_1)\alpha_{t_1}U(t_1,t_2)\alpha_{t_2}...U(t_{n-1},t_n)\alpha_{t_n}U(t_n,t_0)
\end{equation}
where $\alpha_{t_i}$ is a single-time projection operator
representing a property of the system at time $t_i$, and
$U(t,t')=e^{-\frac{i}{\hbar}H(t-t')}$ is the unitary time
evolution operator. \cite{4,5,17,18}

In the `History Projection Operator' (HPO) approach
developed by Isham and collaborators \cite{1,2,3,14}, a
homogeneous history ``$\alpha_{t_{1}}$is true at time
$t_{1}$ and $\alpha_{t_{2}}$ is true at time $t_{2}$ ...
and $\alpha_{t_{n}}$ is true at time $t_{n}$'' is
represented by a {\em projection operator\/} $\alpha$,
defined as the tensor product of projection operators
$\alpha:= \alpha_{t_{1}} \otimes \alpha_{t_{2}} \otimes ...
\otimes \alpha_{t_{n}}$ on the n-fold tensor product of
copies of the standard Hilbert space ${\cal{V}}_{n}:=
{\cal{H}}_{t_{1}}\otimes {\cal{H}}_{t_{2}}\otimes ...
\otimes {\cal{H}}_{t_{n}}$. This approach re-establishes
the logical nature of propositions about a physical system
since these projection operators (and their disjunctions)
represent a type of {\em temporal\/} quantum logic.

Most discussions of the consistent-histories formalism have
involved histories defined at a finite set of time points.
However, it is important to extend this to include a
continuous time variable (especially for potential
applications to quantum field theory), and in order to
construct continuous-time histories on the continuous
tensor product of copies of the Hilbert space
${\cal{V}}_{cts}:= \otimes {\cal{H}}_{t}$, Isham and Linden
defined the History Group \cite{2} as an analogue of the
canonical group of normal quantum theory. This group plays
a crucial role in the physical interpretation of the
theory: the spectral projectors of the generators of its
Lie algebra represent history propositions about the
system.

For the example of a non-relativistic point particle moving
on a line, the history group was defined as a generalised
Weyl group with Lie algebra
\begin{eqnarray}
    [\,x_{t} , x_{t^{\prime}}\,] &=& 0  \\
    {[}\, p_{t} , p_{t^{\prime}}\,] &=& 0  \\
    {[}\, x_{t} , p_{t^{\prime}}\,] &=&
        i \hbar\tau \delta(t -t^{\prime})
\end{eqnarray}
where $-\infty < t, t^{\prime}<+\infty$ and $\tau$ is a
constant with dimensions of time. It is important to
emphasise that the generators of the history algebra $x_t$
and $p_t$, $t\in{\mathbf{R}}$, are Schr\"odinger-picture
operators. After being properly smeared, they correspond
(actually, their spectral projectors), to propositions
about the time-averaged values of the position and the
momentum of the system respectively. The evident
resemblance of the history algebra to the algebra of a
quantum field theory meant that the one-dimensional quantum
mechanics history theory could be treated mathematically in
some respects as a $1+1$ dimension quantum {\em field\/}
theory.

In a previous paper \cite{1}, the requirement of the
existence of the Hamiltonian operator $H$ which represents
propositions about the time-averaged values of the energy
of the system---in particular, for the example of a simple
harmonic oscillator in one dimension---together with the
explicit relation between the Hamiltonian and the creation
and annihilation operators, selected uniquely a Fock space
as the representation space of the history algebra
[1.2-1.4] on the history space ${\cal{V}}_n$. We shall
return to this representation in more detail shortly.

The history algebra generators $x_t$ and $p_t$ can be seen
heuristically as operators, (actually they are
operator-valued distributions on ${\cal{V}}_n$), that for
each time label $t$, are defined on the Hilbert space
${\cal{H}}_t$. The question then arises if, and how, these
Schr\"odinger-picture objects with different time labels
are related: in particular, is there a transformation law
`from one Hilbert space to another'? One anticipates that
the analogue of this question in the context of a histories
treatment of a relativistic quantum field theory would be
crucial to showing the Poincar\'e invariance of the system.

In the Hamilton-Jacobi formulation of Classical Mechanics,
it is the {\em action functional\/} that plays the role of
the generator of a canonical transformation of the system
from one time to another. Indeed, the Hamilton-Jacobi
functional $S$, evaluated for the realised path of the
system---{\em i.e.}, for a solution of the classical
equations of motion, under some initial conditions---is the
generating function of a canonical transformation, which
transforms the system variables position $x$ and momentum
$p$ from an initial time $t=0$ to another time $t$. It is
therefore natural to investigate whether a quantum analogue
of the action functional exists for the HPO theory.

Indeed, in \cite{1} where we explored the quantum field
theory case for the continuous-time histories, we were not
able to show the manifest covariance of the theory under
the `external' Poincar\'e group. However, we did not
consider the action as an operator; the main goal of
the present paper is to enhance the theory in this
direction so as to have a clearer view of the
time-transformation issue. This will ultimately allow us to
re-address the problem of the Poincar\'e covariance of the
quantum field theory \cite{In preparation}

In what follows, we first prove the existence of the action
operator $S_{\kappa}$, using the same type of quantum field
theory methods that were used to prove the existence of the
Hamiltonian operator $H_{\kappa}$ \cite{1}. We will show
that, constructed as a quantum analogue of the classical
action functional, $S_{\kappa}$ does indeed act as a
generator of time-transformations in the HPO theory.
Furthermore---and more speculatively---this is arguably
related to the two laws of time-evolution in standard
quantum theory: namely, wave-packet reduction and the
unitary time-evolution between measurements.

A comparison with the classical theory case seems
appropriate at this point, and thus, in Sections 3 and 4,
we present a classical analogue of the HPO, where the
continuous-time classical histories can be seen to be an
analogue of the continuous-time quantum histories.

In Section 5, we further exploit the classical analogy to
discuss the `classical' behaviour of the history quantum
scheme. In particular, we expect the action operator to be
involved in some way with the dynamics of the theory. To
this end, we show how it appears in the expression for the
decoherence functional expression, with operators acting on
coherent states, as used earlier by Isham and Linden
\cite{2}.

\section{The action operator defined}
In the generalised consistent histories theory by Gell-Mann
and Hartle \cite{4,5} and others, a homogeneous history
$\alpha$ is a time-ordered sequence of propositions about
the system, and is represented by a class operator
$\tilde{C}$
\begin{equation}
\tilde{C}:=
U(t_0,t_1)\alpha_{t_1}U(t_1,t_2)\alpha_{t_2}...U(t_{n-1},t_n)\alpha_{t_n}U(t_n,t_0)
\end{equation}
where $\alpha_{t_i}$ is a single-time projection operator
representing a proposition about the system at time $t_i$.
If a particular history $\alpha$ belongs to a consistent
set, then the probability for the history to be realised is
\begin{equation}
{\rm Prob}(\alpha)=
    tr_{H}{\tilde{C}}^{\dagger}_{\alpha}\rho_{t_{0}}
        {\tilde{C}}_{\alpha}
\end{equation}
where $\rho_{t_{0}}$ is the density matrix of the initial
state. Deciding whether or not a particular set of
histories is consistent involves evaluating the decoherence
functional
\begin{equation}
 d ( \alpha , \beta ) := tr_{H}{\tilde{C}}^{\dagger}_{\alpha}\rho_{t_{0}}{\tilde{C}}_{\beta}
\end{equation}
which is a complex-valued function of a pair of histories
$\alpha$ and $\beta$. In particular, if $\alpha$ and
$\beta$ are disjoint propositions belonging to a consistent
set, then they satisfy the `decoherence' condition
\begin{equation}
 d ( \alpha , \beta ) = 0
\end{equation}

We note that, as a product of projectors, the class
operator $\tilde{C}_{\alpha}$ is generally not itself a
projector, and hence the temporal logic structure of
quantum mechanics is lost. This is remedied in the HPO
theory, in which the history proposition ``$\alpha_{t_{1}}$
is true at time $t_{1}$ and $\alpha_{t_{2}}$ is true at
time $t_{2}$ ... and $\alpha_{t_{n}}$ is true at time
$t_{n}$'' is represented by the tensor product $\alpha:=
\alpha_{t_{1}} \otimes \alpha_{t_{2}} \otimes ... \otimes
\alpha_{t_{n}}$ \cite{14}. This is a genuine projection
operator on the n-fold tensor product ${\cal{V}}_{n}:=
{\cal{H}}_{t_{1}}\otimes {\cal{H}}_{t_{2}}\otimes ...
\otimes {\cal{H}}_{t_{n}}$. Each constituent proposition
$\alpha_t$, labelled by the time parameter t, is defined on a copy of the standard quantum theory Hilbert space, with the
same $t$-label ${\cal{H}}_{t}$.

This is a straightforward idea for a discrete set of times
$(t_1,t_2,\ldots, t_n)$ but, for reasons given in the
Introduction, it is important to extend these ideas to
continuous-time histories which are to be defined on some
sort of continuous tensor product of copies of the Hilbert
space ${\cal{V}}_{cts}:= \otimes {\cal{H}}_{t}$.

A key technical tool is the history group, constructed as
an analogue of the canonical group \cite{17} of normal
quantum theory. For a particle moving in one dimension, the
standard canonical commutation relation
\begin{equation}
   [\,x, p\,] = i \hbar
\end{equation}
is replaced by the `history algebra'
\begin{eqnarray}
    [\, x_{t} , x_{t^{\prime}}\,] &=& 0   \label{W1}   \\
    {[}\, p_{t} , p_{t^{\prime}}\,] &=& 0  \label{W2} \\
    {[}\, x_{t} , p_{t^{\prime}}\,] &=& i
                     \hbar\tau\delta(t-t^{\prime}) \label{W3}
\end{eqnarray}
where $-\infty < t, t^{\prime}<+\infty$. The constant
$\tau$ has dimensions of time \cite{16} and, in what
follows, for convenience we shall choose units in which
$\tau = 1$ . These operators are written in the
Schr\"odinger picture: $t$ labels the Hilbert space---it is
{\em not\/} the time parameter that appears in the
Heisenberg picture for normal quantum theory.

To be mathematically precise, the eqs.\ (\ref{W1}---\ref{W2}) must be smeared
\begin{eqnarray}
      [\,x_{f}, x_{g}\,] & =& 0         \\
     {[}\,p_{f}, p_{g}\,] & =& 0            \\
     {[}\,x_{f}, p_{g}\,] & =& i\hbar\int^{+\infty}_{-\infty}f(t)g(t)dt
\end{eqnarray}
where f and g belong to some appropriate subset of the
space ${\cal{L}}^{2}({\mathbf{R}} , dt)$ of square
integrable functions on ${\mathbf{R}}$.
\par
The evident resemblance of the above with the canonical
commutation algebra of a quantum {\em field\/} theory in
$1+1$ dimensions, leads to the treatment of the history
algebra using mathematical ideas drawn from the former. In
particular, a unique representation of the history algebra
can be selected by the requirement that a representation of
the (analogue of the) Hamiltonian operator exists
\cite{13}: physically, this operator represents history
propositions about the time-averaged values of the energy.\

In previous work \cite{1}, we explored the familiar example
of a simple harmonic oscillator in one dimension. In this
case, the history algebra is extended to include the
commutators
\begin{eqnarray}
    [\,H_{\kappa}, x_{f}\,] & =&
        -\frac{i\hbar}{m}p_{\kappa f}  \label{ha1}   \\
    {[}\,H_{\kappa}, p_{f}\,] & =&
        i\hbar {\omega}^{2}x_{\kappa f} \label{ha2}   \\
  {[}\,H_{\kappa}, H_{\kappa^{\prime}}\,] & =& 0  \label{ha3}
\end{eqnarray}
where $H_{\kappa}$ is the time-averaged history energy
operator $H_{\kappa}:= \int_{-\infty}^{\infty} \kappa (t)
H_{t} dt$. The smearing function $\kappa (t)$ belongs to
some subset of the space ${\cal{L}}^{2}({\mathbf{R}} ,
dt)$, in general not the same as the subset on which the
test functions of the $x_{t}$ and $p_{t}$ are defined. The
specific choice of test functions is partly determined by
the physical situations to which the formalism is to be
applied.

The Fock representation of the history algebra, is based on
the definition of the `annihilation' operator
\begin{equation}
   b_{t}:=\sqrt{\frac{m\omega}{2\hbar}}x_{t}+
        i\sqrt{\frac{1}{2m\omega\hbar}}p_{t}
\end{equation}
with commutation relations
\begin{eqnarray}
      [\,b_{t} , b_{t^{\prime}}\,] &=& 0    \\
     {[}\, b_{t} , b^{\dagger}_{t^{\prime}}\,]
            &=& \delta (t-t^{\prime})
\end{eqnarray}
and is uniquely selected by the requirement that the
time-averaged Hamiltonian operator exists in this
representation; heuristically, $H_t$ is connected with the
operator $b^{\dagger}$ by the expression $ H_{t} =
\hbar\omega b^{\dagger}_{t}b_{t}$.

In the Hamiltonian formalism for a classical system, the
action functional is defined as
\begin{equation}
  S_{\rm{cl}}:=  \int^{+\infty}_{-\infty}(p\dot{q}-H)dt  \label{klas}
\end{equation}
where $q$ is the position, $p$ is the momentum and $H$ the
Hamiltonian of the system. Following the same line of
thought as for the definition of the Hamiltonian algebra,
we want to find a representation of the history algebra in
which their exists a one-parameter family of operators
$S_{t}$---or better their smeared form $S_{\lambda
,\kappa}$. Heuristically we have
\begin{eqnarray}
   S_{t}&:=&(p_{t}\dot{x}_{t}- H_{t})     \\
   S_{\lambda,\kappa}&:=&\int^{+\infty}_{-\infty}
            (\lambda(t)p_{t}\dot{x}_{t}-\kappa(t)H_{t})dt
\end{eqnarray}
where $ S_{\lambda,\kappa} $ is the smeared action operator
with smearing functions $\lambda(t), \kappa(t)$. In order
to discuss the existence of an operator
$S_{\lambda,\kappa}$ we note that, {\em if\/} this operator
exists, the Hamiltonian algebra eqs.\ (\ref{ha1} --- \ref{ha3}), would be augmented in the form
\begin{eqnarray}
[\,S_{\lambda,\kappa}, x_{f}\,]&=&
        i\hbar({x}_{\frac{d}{dt}(\lambda f)} +
            \frac{p_{\kappa f}}{m})  \label{S1}   \\
{[}\, S_{\lambda,\kappa}, p_{f}\,]&=& i\hbar({p}_{\lambda
f} + m\omega x_{\kappa f})  \label{S2}   \\
 {[}S_{\lambda,\kappa} ,
H_{{\kappa}^{\prime}}\,]&=& i\hbar H_{\frac{d}{dt}(\lambda
\kappa^{\prime})} - \frac{i\hbar}{m}
\int^{\infty}_{-\infty} (\kappa^{\prime}(t) \dot{\lambda}
(t)\dot{{p_{t}^{2}}})dt   \label{S3}   \\
 {[}\,S_{\lambda,\kappa} , S_{\lambda,\kappa}^{\prime}\,]&=&
i\hbar H_{\frac{d}{dt}( \lambda^{\prime} \kappa )} - i\hbar
H_{\frac{d}{dt}( \lambda \kappa^{\prime} ) } -\nonumber\\
\hspace{2cm}&&
 i\hbar \int^{\infty}_{-\infty}( [(\kappa(t)
\dot{\lambda^{\prime}}(t))-( \kappa^{\prime}(t)
\dot{\lambda}(t))]\frac{\dot{p_{t}^{2}}}{m}) dt  \label{S4}
\end{eqnarray}

Although we have defined the action operator in a general
smeared form, in what follows we will mainly employ only
the case $\lambda(t)=1$ and $\kappa(t)=1$ that accords with
the expression for the classical action functional. This
choice of smearing functions poses no technical problems
restrictions, provided we keep to the requirement that the
smearing functions for the position and momentum operators
are square-integrable functions. In particular, the
products of the smearing functions $f$ and $g$ in
eqs (\ref{S1}---\ref{S4}) with the test functions $\lambda(t)=1$ and
$\kappa(t)=1$ with, are still square-integrable.

\paragraph*{The existence of the action operator in HPO.}
We now examine whether the action operator actually exists
in the Fock representation of the history algebra employed
in our earlier work\cite{1}. Henceforward we choose
$\lambda(t)=1$. Then the formal commutation relations are
\begin{eqnarray}
 S_{\kappa}&:=& \int^{+\infty}_{-\infty}( p_{t}\dot{x}_{t}- \kappa(t)H_{t})dt     \\
  {[}\, S_{\kappa} , x_{f}\,]&=& i\hbar({x}_{\dot{f}} + \frac{p_{\kappa f}}{m})     \\
  {[}\, S_{\kappa} , p_{f}\,]&=& i\hbar({p}_{ f} + m\omega
            x_{\kappa f}) \\
   {[}\, S_{\kappa} , H_{{\kappa}^{\prime}}\,] &=& i\hbar
        H_{{\dot{\kappa}}^{\prime}} \\
   {[}\, S_{\kappa} , S_{{\kappa}^{\prime}}\,]&=& i\hbar
H_{\dot{\kappa}} -i\hbar H_{{\dot{\kappa}}^{\prime}}
\end{eqnarray}

A key observation is that {\em if\/} the operators
$e^{\frac{i}{\hbar}S_{\kappa}}$ existed they would produce
the history algebra automorphism
\begin{equation}
 e^{\frac{i}{\hbar}S_{\kappa}} b_{t} e^{-\frac{i}{\hbar}S_{\kappa}}=
e^{-i\omega \int^{t+s}_{t} \kappa(t+s^{\prime}) ds^{\prime}
+ s\frac{d}{dt}}b_{t}
\end{equation}
or, in the more rigorous smeared form
\begin{equation}
 e^{\frac{i}{\hbar}S_{\kappa}} b_{f}
 e^{-\frac{i}{\hbar}S_{\kappa}}= b_{{\Sigma}_{s} f}
\end{equation}
where the unitary operator $\Sigma_s$ is defined on
${\mathcal{L}}^{2}({\mathbf{R}})$ by
\begin{equation}
  ({\Sigma}_{s}\psi)(t):=
        e^{-i\omega \int^{t+s}_{t} \kappa(t+s^{\prime})
                ds^{\prime}} \psi(t+s).\label{Def:Sigma}
\end{equation}

However, an important property of the Fock construction
states that when there exists a unitary operator
$e^{i s A }$ acting on
${\mathcal{L}}^{2}({\mathbf{R}})$, there exists a unitary
operator $\Gamma(e^{i s A })$ that acts on the
exponential Fock space\footnote{A general expression for a Fock space is $ e^{\mathcal{H}} = \oplus^{\infty}_{n=0} (\otimes_{n}\mathcal{H})^{n}$ where $\mathcal{H}$ is called the base Hilbert space of the Fock space $e^{\mathcal{H}}$. }
${\mathcal{F}}({\mathcal{L}}^{2}({\mathbf{R}}))$ in such
way that
\begin{equation}
 \Gamma(e^{i s A }) b^{\dagger}_{f}
{\Gamma(e^{i s A })}^{-1} =
{b^{\dagger}}_{e^{i s A }f}   \label{prop1}
\end{equation}
\\
then the operator $d\Gamma( A )$ on ${\mathcal{F}}({\mathcal{L}}^{2}({\mathbf{R}}))$ can also be defined as 
\begin{equation}
\Gamma(e^{i s A }) = e^{i s d\Gamma( A )}   \label{prop2}
\end{equation}
in terms of $A$, a self-adjoint operator that acts on ${\mathcal{L}}^{2}({\mathbf{R}})$.
In particular, it follows that the representation of the
history algebra on the Fock space
${\mathcal{F}}({\mathcal{L}}^{2}({\mathbf{R}}))$ carries a
(weakly continuous) representation of the one-parameter
family of unitary operators $s\mapsto e^{\frac{i}{\hbar}s
S_{\kappa}} = \Gamma(\Sigma_s)$. Therefore, the generator
${S}_{\kappa}$ also exists on
${\mathcal{F}}({\mathcal{L}}^{2}({\mathbf{R}}))$ and $S =
d\Gamma (-\hbar {\sigma}_{\kappa})$ where
${\sigma}_{\kappa}$ is a self-adjoint operator that acts on
${\mathcal{L}}^{2}({\mathbf{R}})$ and is defined as
\begin{equation}
 {\sigma}_{\kappa} \psi (t):=
    \left(-\omega \kappa (t) - i \frac{d}{dt}\right)\psi(t).
\end{equation}

In what follows, we will restrict our attention to the
particular case $\kappa(t) = 1$ for the simple harmonic
oscillator action operator $S$
 \begin{equation}
   S:= \int^{+\infty} _{-\infty} (p_{t}\dot{x}_{t}-
H_{t})dt. \label{Def:op_S}
 \end{equation}

\paragraph*{The Liouville operator definition.} The first
term of the action operator eq.\ (\ref{Def:op_S}) is
identical to the kinematical part of the classical action
functional eq.\ ( \ref{klas} ). For reasons that will become apparent later, we write $S_{\kappa}$ as the difference between two
operators: the Liouville operator and the Hamiltonian
operator. The Liouville operator is formally written as
\begin{equation}
 V:= \int^{\infty}_{-\infty}(p_{t}\dot{q_{t}}) dt
\end{equation}
where
\begin{equation}
 S_{\kappa} = V - H_{\kappa}
\end{equation}
We prove the existence of $V$ on
${\mathcal{F}}({\mathcal{L}}^{2}({\mathbf{R}}))$ using the
same technique as before. Namely, we can see at once that
the history algebra automorphism
\begin{equation}
 e^{\frac{i}{\hbar}sV} b_{f} e^{ - \frac{i}{\hbar}sV} = b_{B_{s}f}
\end{equation}
is unitarily implementable. Here, the unitary operator
$B_s$, $s\in\mathbf{R}$, acting on
${\mathcal{L}}^{2}({\mathbf{R}})$ is defined by
\begin{equation}
 (B_{s}f)(t):= e^{s\frac{d}{dt}}f(t) = e^{i s D}f(t) = f(t+s)
\end{equation}
where $D:= -i\frac{d}{dt}$. The Liouville operator $V$
has some interesting commutation relations with the
generators of the history algebra:
\begin{eqnarray}
  [\, V , x_{f}\,] &=& -i\hbar x_{\dot{f}}          \\
 {[}\, V, p_{f}\,] &=& -i\hbar p_{\dot{f}}          \\
 {[}\, V,H_{\kappa}\,] &=& -i\hbar H_{\dot{\kappa}} \\
 {[}\, V, S_{\kappa}] &=& i\hbar H_{\dot{\kappa}}   \\
 {[}\, V, H\,] &=& 0 \\
 {[}\, V, S\,] &=& 0
\end{eqnarray}
where we have defined $H:= \int^{\infty}_{-\infty}H_{t}dt$.

We notice that $V$ transforms, for example, $b_{t}$ from
one time $t$---that refers to the Hilbert space
${\mathcal{H}}_{t}$---to another time $t+s$, that refers to
${\mathcal{H}}_{t+s}$. More precisely, $V$ transforms the
support of the operator-valued distribution $b_{t}$ from
$t$ to $t+s$:
\begin{equation}
 e^{\frac{i}{\hbar}sV} b_{f}
        e^{ - \frac{i}{\hbar}sV} = b_{f_s}
\end{equation}
where $f_s(t):=f(s+t)$. We shall return to the significance
of this later.

\paragraph*{The Fourier-transformed `$n$-particle' history
propositions.} An interesting family of history
propositions emerges from the representation space
${\mathcal{F}}[{\cal{L}}^{2}({\mathbf{R}} ,dt)]$, acting on
the $\delta$-function normalised basis of states
$|0\rangle$, $|t_{1}\rangle:= b^{\dagger}_{t_{1}}|0\rangle$
, $|t_{1}, t_{2}\rangle:=
b^{\dagger}_{t_{1}}b^{\dagger}_{t_{2}}|0\rangle $ {\em
etc}; or, in smeared form, $|\phi\rangle:=
b^{\dagger}_{\phi}|0 \rangle $ {\em etc}. The projection
operator $ |t\rangle \langle t| $ corresponds to the
history proposition `there is a unit energy $\hbar\omega$
concentrated at the time point t'. The physical
interpretation for this family of propositions, was deduced
from the action of the Hamiltonian operator on the family
of $|t\rangle$ states:
\begin{eqnarray}
   H_{t}|0\rangle &=& 0                                 \\
 H_{t}|t_{1}\rangle &=& \hbar\omega \delta(t -t_{1})
            |t_{1}\rangle                               \\
  H_{t}|t_{1}, t_{2}\rangle &=& \hbar
    \omega [ \delta(t - t_{1}) +\delta(t - t_{2}) ] |t_{1},
        t_{2}\rangle                                    \\
        &\vdots&                              \nonumber
\end{eqnarray}
To study the behaviour of the $S$ operator, a particularly
useful basis for ${\mathcal{F}}[{\cal{L}}^{2}({\mathbf{R}}
, dt)]$ is the Fourier-transforms of the
$|t\rangle$-states. Indeed, if we consider the Fourier
transformations
\begin{eqnarray}
 |\nu \rangle &=& \int^{+\infty}_{-\infty}e^{i\nu
    t}b^{\dagger}_{t}|0\rangle dt                       \\
  |\nu_{1}, \nu_{2}\rangle &=& \int^{+\infty}_{-\infty}
    e^{i\nu_{1}t_{1}} e^{i\nu_{2}t_{2}} b^{\dagger}_{t_{1}}
        b^{\dagger}_{t_{2}}|0\rangle dt_{1}dt_{2}       \\
 b_{\nu}&=& \int^{+\infty}_{-\infty}e^{i\nu t}b_{t} dt  \\
 b^{\dagger}_{\nu}&=& \int^{+\infty}_{-\infty}e^{-i\nu
        t}b^{\dagger}_{t} dt
\end{eqnarray}
the Fourier transformed $|\nu\rangle $- states are defined
by $|\nu\rangle:= b^{\dagger}_{\nu}|0\rangle,
|\nu_{1},\nu_{2}\rangle:=
b^{\dagger}_{\nu_{1}}b^{\dagger}_{\nu_{2}}|0\rangle $ {\em
etc}. The eigenvectors of the operator $S$ are calculated
to be
 \begin{eqnarray}
   S |0 \rangle &=& 0   \\
   S|\nu\rangle &=& \hbar(\nu-\omega)|\nu\rangle        \\
   S|\nu_{1}, \nu_{2} \rangle &=& \hbar [ (\nu_{1} - \omega) +
    (\nu_{2} - \omega) ] |\nu_{1}, \nu_{2} \rangle      \\
        &\vdots&                                \nonumber
\end{eqnarray}
and we note in particular that
$e^{\frac{i}{\hbar}sS}|0\rangle = |0\rangle$.

The $|\nu \rangle$-states are also eigenstates of the
Hamiltonian operator:
\begin{eqnarray}
   H |0 \rangle &=& 0   \\
   H|\nu\rangle &=& \hbar \omega |\nu\rangle    \\
   H|\nu_{1}, \nu_{2} \rangle &=& 2 \hbar
        \omega |\nu_{1}, \nu_{2} \rangle        \\
        &\vdots&                            \nonumber
\end{eqnarray}
Again, as for the case of the $|t \rangle$- states, for the
special case of $H:= \int^{\infty}_{-\infty} H_{t}dt$, and
for the simple harmonic oscillator example, we see how the
integer-spaced spectrum of the standard quantum field
theory appears in the HPO theory. The
$|\nu\rangle\langle\nu |$ history propositions give the
spectrum of the action operator and they have an
interesting connection with the $|t\rangle\langle t| $
propositions.

\subsection{The Velocity operator}
In \cite{1}, we emphasised the existence of the operator
${\dot{x}}_{t}:= \frac{d}{dt}x_{t}$, that corresponds to
history propositions about the velocity of the system. The
velocity operator is better defined in its smeared form
using the familiar quantum field theory procedure
\begin{equation}
  {\dot{x}}_{f} = -x_{\dot{f}}  \label{Def:dotx}
\end{equation}
In analogy with quantum field theory, this requires the
function $f$ to be differentiable and to `vanishes at
infinity' so that the implicit integration by parts in eq
(\ref{Def:dotx}) is valid. We note that, in this HPO
theory, the velocity operator commutes with the position
\begin{equation}
 [\,x_{t} , {\dot{x}}_{t^{\prime}}\,] = 0,
\end{equation}
and therefore there exist history propositions about the
position and the velocity at the same time.

Furthermore, the existence of the Liouville operator in the
HPO scheme, allows an interesting comparison between the
velocity ${\dot{x}}_{f}$ and the momentum $p_{f}$
operators: namely, the momentum operator is defined by the
history commutation relation of the position with the
Hamiltonian, while we can define the velocity operator from
the history commutation relation of the position with the
Liouville operator:
\begin{eqnarray}
  [\,x_{f}, H\,] &=& i\hbar \frac{p_{f}}{m}   \\
  {[}\, x_{f}, V \,] &=& i\hbar{\dot{x}}_{f}
\end{eqnarray}
These relations signify the different nature of the
momentum $p_{f}$ from the velocity ${\dot{x}}_{f}$
concerning the dynamical behaviour of the momentum (related
to the Hamiltonian operator), as opposed to the kinematical
behaviour of the velocity (related to the Liouville
operator).

\subsection{The Heisenberg picture}
In standard quantum theory, a Heisenberg-picture operator
$A(s)$ is defined as
\begin{equation}
 A_{H}(s):= e^{\frac{i}{\hbar} s H} A e^{ - \frac{i}{\hbar} s H}
\end{equation}
In particular, for the case of a simple harmonic
oscillator, the equation of motion is
\begin{equation}
 \frac{d^{2}}{ds^{2}}x(s)+ \omega^{2}x(s)= 0
\end{equation}
from which we obtain the solution
\begin{eqnarray}
  x(s)= \cos (s\omega ) x + \frac{1}{m\omega} \sin (s\omega ) p \\
  p (s)= - m\omega \sin (s\omega ) x + \cos (s\omega ) p
\end{eqnarray}
where we have used the classical equation
\begin{equation}
p:= m\frac{dx(s)}{ds}\Big |_{s=0}
\end{equation}
The commutation relations between these operators is
\begin{equation}
  [\,x(s_{1}) , x( s_{2})\,] =
    \frac{i\hbar}{m\omega} \sin [\omega(s_{1}-s_{2})]
\end{equation}

In formulating a history analogue of the Heisenberg picture
\cite{1}, we adopted a `time-averaged' Heisenberg picture
defined by
\begin{equation}
    x_{\kappa,t}:=e^{i/ \hbar H_{\kappa}}x_{t}
    e^{-i/\hbar H_{\kappa}}=\cos[\omega\kappa(t)]x_{t}+
        \frac{1}{m\omega}\sin[\omega\kappa(t)]p_{t}\\
\end{equation}
for suitable test functions $\kappa$. The analogue of the
equations of motion is the functional differential equation
\begin{equation}
   \frac{\delta^{2}x_{\kappa,t}}{\delta\kappa(s_{1} )
    \delta\kappa(s_{2})}+\delta(t-s_{1})\delta(t-s_{2})
        \omega^{2}x_{\kappa,t} = 0
\end{equation}
and
\begin{equation}
     \delta(t-s)p_{t} = m\frac{\delta x_{\kappa,t}}{\delta
\kappa(s)}\Big |_{\kappa=0}
\end{equation}
is the history analogue of the classical equation $ p:=
m\frac{dx(s)}{ds}\Big |_{s=0} $.

We noted then that the Heisenberg-picture in an HPO theory
involves two time labels: an `external' label $t$---that
specifies the time the proposition is asserted---and an
`internal' label $s$ that, for a fixed time $t$, is the
time parameter of the Heisenberg picture associated with
the copy ${\mathcal{H}}_{t}$ of the standard Hilbert space.
Using our new results, the two labels appear naturally in a
new version of the Heisenberg picture: they are related to
the groups that produce the two types of time
transformations. In addition, the analogy with the
classical expressions is regained.

To see this explicitly, we define a Heisenberg picture
analogue of $x_{t}$ as
\begin{eqnarray}
x_{\kappa,t,s } :&=& e^{\frac{i}{\hbar} s H_{\kappa}} x_{t}
e^{-\frac{i}{\hbar}sH_{\kappa}} \\
&=& \cos[\omega s \kappa(t)]x_{t} + \frac{1}{m\omega}\sin[\omega s \kappa(t)]p_{t}   \nonumber  \\
p_{ \kappa,t,s } :&=& e^{ \frac{i}{\hbar} s H_{\kappa}}
p_{t} e^{-\frac{i}{\hbar} s H_{\kappa}} \\
    &=& -m \omega \sin [\omega s \kappa(t)]x_{t} + \cos [\omega s \kappa(t)]p_{t} \nonumber
\end{eqnarray}
The commutation relations for these operators are
\begin{eqnarray}
 [\,x_{\kappa,t}(s),x_{\kappa^{\prime},t^{\prime}}(s^{\prime})\,]
        &=& \frac{i\hbar}{m\omega}\sin[\omega\kappa
            (s^{\prime}-s)]\delta(t-t^{\prime})             \\
    {[}\,x_{\kappa,t}(s), S_{{\kappa}^{\prime}}\,] &=&
     i\hbar \Big{[} \cos[s\omega \kappa(t)] \dot{x_{t}}
      + \frac{1}{m\omega} \sin[s\omega \kappa(t)]
        \dot{p_{t}} - \frac{{\kappa}^{\prime}}{m}
            p_{\kappa , t, s} \Big{]}  \hspace{1cm}                     \\
    {[}\,p_{\kappa,t}(s), S_{{\kappa}^{\prime}}\,] &=&
        i\hbar\Big{[}\cos[s\omega \kappa(t)] \dot{p_{t}} -
        m\omega \sin[s\omega \kappa(t)] \dot{x_{t}} +
        {\kappa}^{\prime}(t) x_{\kappa,t,s}\Big{]}\hspace{1cm}
\end{eqnarray}
and from these commutators we obtain the HPO analogue of
the equations of motion
\begin{equation}
 \frac{d^{2}}{ds^{2}}x_{\kappa, t, s}+ \omega^{2} {\kappa(t)}^{2} x_{\kappa, t, s}= 0
\end{equation}
We notice the strong resemblance with standard quantum
theory; for the case $\kappa(t)=1$, the classical
expressions are fully recovered.

In the HPO formalism, the Heisenberg picture objects appear
time averaged with respect to the `external' time label
$t$. On the other hand, the `internal' time label $s$ is
the time-evolution parameter of the standard Heisenberg
picture, as viewed in the Hilbert space
${\mathcal{H}}_{t}$. In what follows, we will show how the
Heisenberg picture operators evolve in time under the
action of the groups of time transformations.

\section{Time transformation in the HPO formalism}
In classical theory, the Hamiltonian $H$ is the generator of
time transformations. In terms of Poisson brackets, the
generalised equation of motion for an arbitrary function $u$ is
given by
\begin{equation}
 \frac{du}{dt} = \{ u , H \} + \frac{\partial u}{\partial t}.
\end{equation}

In a HPO theory, the Hamiltonian operator $H_t$ produces phase
changes in time, preserving the time label $t$ of the Hilbert
space on which, at least formally, $H_t$ is defined. On the
other it is the Liouville operator $V$ that assigns, analogous
to the classical case, history commutation relations, and
produces time transformations `from one Hilbert space to
another'. The action operator generates a combination of these
two types of time-transformation. If we use the notation
$x_{f}(s)$ for the $\it{history}$ Heisenberg-picture operators
smeared with respect to the time label $t$, we observe that they
behave as standard Heisenberg-picture operators, with time
parameter $s$. Furthermore, their history commutation relations
strongly resemble the classical expressions:
\begin{eqnarray}
[\,x_{f}(s) , V\,] &=& i\hbar {\dot{x}}_{f}(s) \\
  {[}\, x_{f}(s), H\,] &=& \frac{i\hbar}{m}p_{f}(s)         \\
  {[}\, x_{f}(s), S\,] &=&
    i\hbar({\dot{x}}_{f}(s)- \frac{1}{m}p_{f}(s))
\end{eqnarray}

 We define a one-parameter group of transformations $T_{V}(\tau)$,
with elements $e^{\frac{i}{\hbar}\tau V}$, $\tau\in{\mathbf{R}}$
where $V$ is the Liouville operator and we consider its action
on the $b_{t}$ operator; for simplicity we write the unsmeared
expressions
\begin{equation}
 e^{\frac{i}{\hbar}\tau V} b_{t,s} e^{ - \frac{i}{\hbar}\tau V} = b_{t+\tau,s}
\end{equation}
The Liouville operator is the generator of transformations of
the time parameter $t$ labelling the Hilbert spaces
${\mathcal{H}}_{t}$.

Then, we define a one-parameter group of transformations
$T_{H}$, with elements $e^{\frac{i}{\hbar}\tau H}$, where $H$ is
the time-averaged Hamiltonian operator
\begin{equation}
 e^{\frac{i}{\hbar}\tau H} b_{t,s} e^{ - \frac{i}{\hbar}\tau H} = b_{t,s+\tau}
\end{equation}
The Hamiltonian operator is the generator of phase changes
of the time parameter $s$, produced only on one Hilbert
space ${\mathcal{H}}_{t}$, for a fixed value of the $t$
parameter.
\par
Finally, we define the one-parameter group of
transformations $T_{S}$, with elements
$e^{\frac{i}{\hbar}\tau S}$, where $S$ is the action
operator
\begin{equation}
 e^{\frac{i}{\hbar}\tau S} b_{t,s} e^{ - \frac{i}{\hbar}\tau S} = b_{t+\tau , s+\tau}
\end{equation}
The action operator generates both types of time
transformations---a feature that appears only in the HPO scheme.

In Fig.1a,b, we denote a quantum continuous-time history as a
curve and the tensor product of Hilbert spaces as a sequence of
planes, each one representing a copy of the standard Hilbert
space. Each plane is labeled by the time label $t$ that the
corresponding Hilbert space $H_t$ carries. We depict then a
history, as a curve along an $n$-fold sequence of `Hilbert
planes' ${\mathcal{H}}_{t_{i}}$. In analogy to this, we
symbolise a classical history, as a curve along an $n$-fold
sequence of planes corresponding to copies of the standard
phase-space $\Gamma_{t_{i}}$, as we will explain later. The time
transformations generated by the Liouville operator, shift the
path in the direction of the `Hilbert planes'. On the other
hand, the Hamiltonian operator generates time transformations
that move the history curve in the direction of the path, as
represented on one `Hilbert plane'.
\begin{figure}[h]
\caption{Quantum
and classical history curves. In Figure 1.a the transformation
of the history curves generated by $V$ is represented by the
dashed line, while the transformation generated by $H$ are
represented by the dotted line. The curves drawn on each `Hilbert plane' correspond to the Hamiltonian transformations as effected on
the corresponding Hilbert space. In Figure 1.b the classical history remains invariant under the corresponding time transformations }
\label{fig:test}
\end{figure}

\paragraph*{The duality in time-evolution} In standard quantum
theory, time-evolution is described by two different laws: the
wave-packet reduction that occurs at a measurement, and the
unitary time-evolution that takes place between measurements.
Thus, according to von Neumann, one has to augment the
Schr\"odinger equation with the collapse of the wave-function
associated with a measurement \cite{9}.

It seems that the two types of time-transformations observed in
the HPO theory, correspond to the two dynamical processes in
standard quantum theory: the time transformations generated by
the Liouville operator $V$ are (arguably) related to the
wave-packet reduction (the time ordering implied by the
wave-packet reduction to be precise), while the time
transformations produced by the Hamiltonian operator $H$ are
related to the unitary time-evolution between measurements.

The argument in support of this assertion is as follows. Keeping
in mind the description of the History space as a tensor product
of single-time Hilbert spaces ${\mathcal{H}}_{t}$, the $V$
operator acts on the Schr\"odinger-picture projection operators
translating them in time from one Hilbert space to another.
These time-ordered projectors appear in the expression for the
decoherence functional that defines probabilities. In history
theory, the expression for probabilities in a consistent set is
the same as that derived in standard quantum theory using the
projection postulate on a time-ordered sequence of measurements
\cite{4,5}. It is this that suggests a relation of the Liouville
operator to `wave-packet reduction'. To strengthen this claim,
in what follows we will show the analogy of $V$ with the
$S_{cts}$ operator (an approximation of the derivative
operator), that appears in the decoherence function and is
implicitly related to the wave-packet reduction by specifying
the time-ordering of the action of the single-time projectors.
The action of $V$ as a generator of time translations depends on
the partial (in fact, total) ordering of the time parameter
treated as the causal structure in the underlying spacetime.
Hence, the $V$-time translations illustrate the purely
kinematical function of the Liouville operator.

The Hamiltonian operator producing transformations, with an
evident reference to the Heisenberg time-evolution, appears as
the `clock' of the theory. As such, it depends on the particular
physical system that the Hamiltonian describes. Indeed, we would
expect the definition of a `clock' for the evolution in time of
a physical system to be connected with the dynamics of the
system concerned. We note that the idea of reparametrizing time
depends on the smearing function $\kappa (t)$ used in the
definition of the Hamiltonian operator; $\kappa (t)$ is kept
fixed for a particular physical system.

The coexistence of the two types of time-evolution, as reflected
in the action operator identified as the generator of such time
transformations, is a striking result. In particular, its
definition is in accord with its classical analogue, namely the
Hamilton action functional. In classical theory, a distinction
between a kinematical and a dynamical part of the action
functional also arises in the sense that the first part
corresponds to the symplectic structure and the second to the
Hamiltonian.

\section{The classical signature of the HPO formalism}

Let us now consider more closely the relation of the classical
and the quantum histories. We have shown above how the action
operator generates time translations from one Hilbert space to
another, through the Liouville operator; and on each labeled
Hilbert space ${\cal{H}}_{t}$, through the Hamiltonian operator.
We now wish to discuss in more detail the analogue of these
transformations in the classical case.

We recall that a history is a time-ordered sequence of
propositions about the system. The continuous-time quantum
history in the HPO system, makes assertions about the values of
the position or the momentum of the system, or a linear
combination of them, at each moment of time, and is represented
by a projection operator on the continuous tensor product of
copies of the standard Hilbert space.

One expects that a continuous-time classical history should
reflect the underlying temporal logic of the situation. Thus the
assertions about the position and the momentum of the system at
each moment of time should be represented on an analogous
history space: this can be achieved by using the Cartesian
product of a continuous family (labelled by the time $t$) of
copies of the standard classical state space.

In classical mechanics, a (fine-grained) classical history is
represented by a path in the state space. Indeed, a path
$\gamma$ is defined as a map on the standard phase-space
\begin{eqnarray}
     \gamma : & {\mathbf{R}}& \to \Gamma    \\
     & t& \mapsto (q(\gamma(t)), p(\gamma(t)))    \nonumber
\end{eqnarray}
where $q((\gamma(t))$ and $p(\gamma(t))$ are the position and
momentum coordinates of the path $\gamma$, at the time $t$. For
our purposes, we shall consider the path $t\mapsto\gamma(t)$ to
defined for $t$ in some finite time interval $[t_1, t_2]$. We
shall denote the set of such paths by $\Pi$.

The key idea of this new approach to classical histories is
contained in the symplectic structure of the theory: the choice
of the Poisson bracket must be such that it includes entries at
different moments of time. Thus we suppose that the space of
functions on $\Pi$ is equipped with the `history Poisson
bracket' defined by
\begin{equation}
  \{q_t , p_{t^{\prime}}\}= \delta(t-t^{\prime})
\end{equation}
where we defined the functions $q_t$ on $\Pi$ as
\begin{eqnarray*}
            q_t : &\Pi& \to {\mathbf{R}} \\
                  &\gamma& \mapsto q_t(\gamma):= q(\gamma(t))
\end{eqnarray*}
and similarly for $p_t$.

 We now define the history action functional $S_{h}(\gamma)$ on
 $\Pi$ as
\begin{equation}
   S_{h}(\gamma):=
   \int^{t_2}_{t_1}{[p_t\dot{q_t}-H_t(p_t,q_t)](\gamma)\,dt}
\end{equation}
where $q_t(\gamma)$ is the position coordinate $q$ at the time
point $t \in [t_1, t_2]$ of the path $\gamma$, and
$\dot{q_t}(\gamma)$ is the velocity coordinate at the time point
$t\in[t_1, t_2]$ of the path $\gamma$.

We also define the history classical analogues for the Liouville
and time-averaged Hamiltonian operators as
\begin{eqnarray}
 V_{h} ( \gamma )&:=& \int^{t_2}_{t_1}{[p_t\dot{q_t}](\gamma)\,dt}    \\
 H_{h} ( \gamma )&:=& \int^{t_2}_{t_1}{[H_t(p_t,q_t)](\gamma)\,dt}    \\
   S_{h}( \gamma ) &=& V_{h}( \gamma )- H_{h}( \gamma )
\end{eqnarray}

In classical mechanics, the least action principle states that,
there exists a functional $S(\gamma )=
\int^{t_2}_{t_1}{[p\dot{q}-H(p,q)](\gamma)\,dt}$ such that the
physically realised path is the  curve in state space,
${\gamma}_{0}$, with respect to which the condition $\delta S
({\gamma}_{0}) = 0$ holds, when we consider variations around
this curve. From this, the Hamilton equations are deduced to be
\begin{eqnarray}
  \dot{q} &=& \{ q , H \}         \\
  \dot{p} &=& \{ p , H \}
\end{eqnarray}
where $q$ and $p$---the coordinates of the realised path
${\gamma}_{0}$---are the solutions of the classical equations of
motion. For any function $F(q,p)$ of the classical solutions it
is also true that
\begin{equation}
    \{F, H \} = \dot{F}
\end{equation}

In the case of the classical continuous-time histories, one can
formulate the above variational principal in terms of the
Hamilton equations with the statement: A classical history
${\gamma}_{cl}$ is the realised path of the system---{\em
i.e.\/} a solution of the equations of motion of the system---if
it satisfies the equations
\begin{eqnarray}
     \{q_t , V_{h}\}(\gamma_{cl}) = \{q_t , H_{h}\}(\gamma_{cl}) \label{Ham1}    \\      
      \{p_t , V_{h}\}(\gamma_{cl}) = \{p_t , H_{h}\}(\gamma_{cl}) \label{Ham2}
\end{eqnarray}
where ${\gamma}_{cl}= (q_t({\gamma}_{cl}) , p_t({\gamma}_{cl})
)$,  and  $q_t({\gamma}_{cl})$ is the position coordinate of the
realised path ${\gamma}_{cl}$ at the time point $t$. The eqs.\
(\ref{Ham1} -- \ref{Ham2}) are the history equivalent of the Hamilton
equations of motion. Indeed, for the case of the simple harmonic
oscillator in one dimension the eqs.\ (\ref{Ham1} -- \ref{Ham2}) become
\begin{eqnarray}
     {\dot{q}}_t ({\gamma}_{cl}) &=& \frac{p_t}{m}({\gamma}_{cl})   \\
     {\dot{p}}_t ({\gamma}_{cl}) &=& - m {\omega}^{2} q_t ({\gamma}_{cl})
\end{eqnarray}
where ${\dot{q}}_t ({\gamma}_{cl}) = \dot{q}({\gamma}_{cl}(t))|$ is
the value of the velocity of the system at time $t$. One would
have expected the result in eqs.\
(\ref{Ham1}---\ref{Ham2}) for the classical
analogue of the histories formalism, as it shows that the
classical analogue of the two types of time-transformation in
the quantum theory coincide.

 From the eqs.\
(\ref{Ham1}---\ref{Ham2}) we also conclude
that the canonical transformation generated by the history
action functional $S_{h}({\gamma}_{cl})$, leaves invariant the
paths that are classical solutions of the system:
\begin{eqnarray}
     \{q_t , S_{h} \}(\gamma_{cl})&=& 0  \label{clsol1} \\
     \{p_t , S_{h} \}(\gamma_{cl})&=& 0  \label{clsol2}
\end{eqnarray}
It also holds that any function $F$ on $\Pi$ satisfies the
equation
\begin{equation}
     \{ F, S_{h}\} (\gamma_{cl})= 0
\end{equation}
\par
Some of these statements are implicit in previous work by C.
Anastopoulos \cite{6}; an interesting application of a similar
extended Poisson bracket using a different formulation has been
done by I.Kouletsis \cite{7}.

\paragraph*{`Classical' coherent states for the simple harmonic oscillator.}
The relation between the classical and the quantum theories can
be further exemplified by using  coherent states. This special
class of states was used in \cite{2} to represent certain
continuous-time history propositions in the history space.
Coherent states are particularly useful for this purpose since
they form a natural (over-complete) base for the Fock space
representation of the history algebra.

A class of coherent states in the relevant Fock space is
generated by unitary transformations on the cyclic vacuum state:
\begin{equation}
          |f,h\rangle := U[f,h]|0\rangle
\end{equation}
where $ U [f,h]$ is the Weyl operator defined as
 \begin{equation}
    U [f,h]:= e^{\frac{i}{\hbar}(x_f- p_h)},
  \end{equation}
where $f$ and $h$ are test functions in $L^2({\mathbf{R}})$. The
Weyl generator
 \begin{equation}
      \alpha (f,h):= x (f)- p (h)
 \end{equation}
 can alternatively be written as
 \begin{equation}
 \alpha (f,h)= \frac{\hbar}{i}(b^{\dagger}(w)-b(w^{*}))
 \end{equation}
where $w:=f+ih$.

 Suppose now that, for a pair of functions $(f,h)$, the
operator $\alpha(f,h)$ commutes with the action operator $S$
\begin{equation}
   [\,S, \alpha(f,h)\,]= 0     \label{alpha(f,h)}
\end{equation}
Then any pair $(f,h)$ satisfying this equation is necessarily a
solution of the system of differential equations obtained from
eq.\ (\ref{alpha(f,h)}):
\begin{eqnarray}
              \dot{f}+ m\omega^2h &=& 0 \label{system1} \\
              \dot{h}- \frac{f}{m} &=& 0   \label{system2}
\end{eqnarray}
We see that if we identify $f$ with the classical momentum
$p_{cl}$ and $h$ with the classical position $x_{cl}$, then the
eqs.\ (\ref{system1}-- \ref{system2}) are precisely the classical equations of
motion for the simple harmonic oscillator:
\begin{equation}
 \ddot{x_{cl}} + \omega^2 x_{cl}= 0.
\end{equation}

The classical solutions $(f,h)$ distinguish a special class of
Weyl operators ${\alpha}_{cl} (f,h)$, and hence a special class
of coherent states:
\begin{equation}
  | \exp z_{cl} \rangle:= U_{{\alpha}_{cl} (f,h)}|0 \rangle
\end{equation}
where $z_{{\rm cl}}:=f+ih$.

These classical-like features stem from the following relation
with $S$
\begin{equation}
  [\,S , U_{{\alpha}_{cl}}\,] = 0
\end{equation}
\begin{equation}
  [\,S , P_{|\exp z _{cl}\rangle}\,] = 0  \label{acpr}
\end{equation}
where $ P_{|\exp z_{cl} \rangle}$ is the projection operator
onto the (non-normalised) coherent state $|\exp z_{cl}\rangle $:
\begin{equation}
   P_{|\exp z_{cl}\rangle}:= \frac{|\exp z_{cl}\rangle
   \langle \exp z_{cl} |}{\langle \exp z_{cl}|\exp
   z_{cl}\rangle}   \label{proj}
\end{equation}

We note that there exists an analogy between the eqs.\
(\ref{clsol1}-- \ref{clsol2}) and the eq.\ (\ref{acpr}), if we consider $(f,h)$ to be
the classical solution: $t\mapsto (q_t, p_t)(\gamma_{cl})$. In
classical histories, the canonical transformation eqs.\
(\ref{clsol1}-- \ref{clsol2}) generated by the history action functional vanishes
on a solution to the equations of motion. On the other hand,
when we deal with quantum histories, the action operator
produces the classical equations of motion eqs.\ (\ref{system2}-- \ref{system1})
when we require that it commutes with the projector (as in \
eq.\ (\ref{proj})) which corresponds to a classical solution $(f,h)$ of the
system. However, we do not imply from this the appearance of the
classical limit: to make any such physical predictions we must
involve the decoherence functional and the coarse graining
operation.

Notice that the construction above holds for a generic
potential, as long as there exists a representation on $
{\cal{V}}_{cts}$ of the history algebra on which the action
operator is defined.

\section{The decoherence functional argument}
In the consistent histories quantum theory, the dynamics of a
system is described by the decoherence functional. In a
classical theory it is the action functional that plays a
similar role in regard to the dynamics of the system. It is only
natural then, to seek for the appearance of the action operator
in the decoherence functional. The aim is to write the HPO
expression for the decoherence functional, with respect to an
operator that includes $S$, and to compare this operator ({\em
i.e.}, its matrix elements), with the operator
${\mathcal{S}}_{cts} \mathcal{U}$ that appears in the
decoherence functional \cite{2}.

In the HPO formalism, the decoherence functional $d$ has been
constructed for the special case of continuous-time projection
operators corresponding to coherent states \cite{2}. To this
end, a continuous product of projectors $\otimes_{t}
P_{|\exp\lambda(t)\rangle}$ is identified with $ P_{ \otimes_{t} |\exp
\lambda(\cdot)\rangle}$: the projector onto the (non-normalised)
coherent states $\otimes_{t}|\exp \lambda(t)\rangle$ in the
continuous tensor product $\otimes_{t}L^{2}_{t}(\mathcal{R})$.
More precisely, this continuous tensor product is isomorphic to
Fock space:
\begin{equation}
 {\mathcal{V}}_{cts}:=  {\otimes}_{t\in{\mathbf{R}}}{{\cal{L}}^{2}}_{t}({\mathbf{R}})\approx \exp({\cal{L}}^{2}({\mathbf{R}},dt))
\end{equation}
and we can identify a projector on the Hilbert space
${\mathcal{V}}_{cts}$ as
\begin{equation}
 {\otimes}_{t\in{\mathbf{R}}} P_{|\exp(\lambda(t)\rangle}
 = P_{|\exp \lambda(\cdot)\rangle}
\end{equation}
with $ P_{|\exp \lambda(\cdot)\rangle} = e^{-\langle\lambda,
\lambda\rangle} |\exp \lambda(\cdot) \rangle\langle \exp
\lambda(\cdot)| $. The action of the continuous-time histories
projectors on the non-normalised coherent states $
|\exp(\lambda(\cdot))\rangle $ is denoted by
\begin{equation}
 P_{|\exp \lambda(\cdot)\rangle}|\exp(\mu(\cdot))\rangle = e^{-\langle\lambda,  \mu -\lambda\rangle}|\exp \lambda(\cdot)\rangle
\end{equation}

The decoherence functional $d(\mu,\nu)$ for two continuous-time
histories is denoted by
\begin{equation}
  d(\mu,\nu) = tr_{{\mathcal{V}}_{cts}\otimes
  {\mathcal{V}}_{cts}}( P_{|\exp \mu(\cdot)\rangle}
  \otimes P_{|\exp (\nu(\cdot)\rangle} X)
  \end{equation}
where
 \begin{equation}
  X:= \langle 0 | \rho_{-\infty}| 0
  \rangle({\mathcal{S}}_{cts}{\mathcal{U}})^{\dagger}
  \otimes({\mathcal{S}}_{cts}{\mathcal{U}}).
\end{equation}
The operator ${\mathcal{S}}_{{\rm cts}}$ that appears in this
expression for the $d(\mu,\nu)$ was defined as an approximation
of the derivative operator in the sense that
\begin{equation}
  {\mathcal{S}}_{cts}|\exp \nu(\cdot)\rangle =
    |\exp(\nu(\cdot) + \dot{\nu}(\cdot))\rangle   \label{Scts}
\end{equation}
while the dynamics was introduced by the operator ${\mathcal{U}}$, defined in such way that the notion of time evolution is encoded
\begin{equation}
  e^{\langle \lambda, \dot{\lambda} \rangle}
  e^{\frac{i}{\hbar}H[\lambda]} =
  tr_{{\mathcal{V}}_{cts}}({\mathcal{S}}_{cts}
  {\mathcal{U}}{\cal{P}}_{|\exp \lambda(\cdot)\rangle}) \label{U}
\end{equation}

We expect $V$ and $H$ to play a similar role to that of
${\mathcal{S}}_{cts}$ and ${\mathcal{U}}$ respectively, inside
an expression for the decoherence functional. To demonstrate
this we will use the type of Fock space construction given in
eqs.\ (\ref{prop1}---\ref{prop2}). In particular, we use the property
\begin{equation}
 \Gamma(A) |\exp\nu(\cdot)\rangle = |\exp(A\nu(\cdot))\rangle
\end{equation}
where $A$ is an operator that acts on the elements $ \nu
(\cdot)$ of the base Hilbert space $\mathcal{H}$, while the
operator $\Gamma(A)$, defined by eq.\ (\ref{prop1}), acts on the
coherent states $|\exp\nu(\cdot)\rangle$ of the Fock space
$e^{\mathcal{H}}$.

We notice that ${\mathcal{U}}$ is related to the unitary
time-evolution eq.\ (\ref{U}) in a similar way to that of the
Hamiltonian operator $H$
\begin{equation}
    e^{isH} |\exp\nu(\cdot) \rangle = \Gamma ( e^{is\omega I})
|\exp\nu(\cdot)\rangle = |\exp(e^{is\omega} \nu(\cdot) )\rangle
    \label{Action_eisH}
\end{equation}
where $I$ is the unit operator. We also notice that the action
of the operator $e^{isH}$ produces phase changes, as reflected
in the right hand side of eq.\ (\ref{Action_eisH}) (which has
been calculated for for the special case of the simple harmonic
oscillator). Furthermore, when the operator
${\mathcal{S}}_{cts}$ acts on a coherent state eq.\ (\ref{Scts}), it transforms it to another coherent state which involves the
addition to the defining function $\nu(\cdot)$ in a way that
involves the time derivative of $\nu$; and it is noteworthy that
the Liouville operator $V$ acts in a similar way:
\begin{equation}
 e^{isV} |\exp\nu(\cdot) \rangle =
 \Gamma ( e^{isD}) |\exp\nu(\cdot)\rangle =
 |\exp(e^{isD}\nu (\cdot))\rangle
 \end{equation}
 where
 \begin{equation}
  (e^{isD} \nu)(t)  = \nu(t+s)
\end{equation}
where $D:= -i\frac{d}{dt}$. The operator $e^{isD}$ acts on the
base Hilbert space, and corresponds to the operator $e^{isV}$
under the $\Gamma$-construction on the Fock space; that is, it
acts on the vector $\nu(t)$ and transforms it to another one $
\nu(t+s)$, which, for each time $t$ is translation by the time
interval $s$.

This suggests that we define the operator $ {\mathcal{A}}_{s} :=
e^{i s S}$, where $S:= \int^{+\infty} _{-\infty}
(p_{t}\dot{x}_{t}- H_{t})dt$ is the action operator for the
simple harmonic oscillator, which one expects to be related to
the operator ${\mathcal{S}}_{cts}{\mathcal{U}}$. For this
reason, we write the matrix elements of both operators and
compare them.

The general formula for the matrix elements of an arbitrary
operator ${\mathcal{T}}$ with respect to the coherent states
basis in the history space that was used in \cite{2} is
\begin{eqnarray}
 \langle
\exp \mu(\cdot)| {\mathcal{T}} |\exp \nu(\cdot)\rangle&=&
e^{(\langle \mu, \frac{\delta}{\delta \bar{\lambda}} \rangle +
\langle \frac{\delta}{\delta \lambda}, \nu \rangle)} \langle
\exp \lambda(\cdot)| {\mathcal{T}} | \exp \lambda(\cdot) \rangle
{\Big{|}}_{\lambda={\bar{\lambda}}=0}
\end{eqnarray}
hence we need only compare the diagonal matrix elements of the
two operators ${\mathcal{S}}_{cts}{\mathcal{U}}$ and ${\mathcal
A}_s$. Thus we have
\begin{equation}
 \langle
\exp(\lambda(\cdot)|{\mathcal{S}}_{cts}{\mathcal{U}}|\exp(\lambda(\cdot)\rangle
= e^{\langle\lambda, \lambda +\dot{\lambda}\rangle} e^{\frac{i}{\hbar}H[\lambda]}
\end{equation}
where $H[\lambda]:= \int^{\infty}_{-\infty}H(\lambda(t))dt $ and
$H(\lambda):=H(\lambda, \lambda )= \langle \lambda | H | \lambda
\rangle / \langle \lambda | \lambda \rangle$; and
\begin{equation}
  \langle \exp\lambda(\cdot)|{\mathcal{A}}_{s}|\exp\lambda(\cdot)\rangle=
e^{\langle \lambda , e^{is ( \omega I + D )} \lambda \rangle}
\end{equation}
with
\begin{equation}
  (e^{is( \omega I +  D )} \lambda)(t) = e^{is\omega} \lambda (t+s)
\end{equation}

We can also write both of the above operators on the history
space ${\mathcal{F}}({\mathcal{L}}^{2}({\mathbf{R}}))$ using
their corresponding operators on the Hilbert space
${\mathcal{L}}^{2}({\mathbf{R}})$. The $\Gamma$ construction
shows that
\begin{eqnarray}
  {\mathcal{S}}_{cts}{\mathcal{U}} &=& \Gamma(1 + i\sigma) \\
  {\mathcal{A}}_{s} &=&\Gamma(e^{is\sigma}) = e^{isd\Gamma(\sigma)}
\end{eqnarray}
where $\sigma=\omega I + iD$, and $I$ is the unit operator. As
expressions of the same function $\sigma $, the operators
${\mathcal{S}}_{{\rm cts}}{\mathcal{U}}$ and $
{\mathcal{A}}_{s}$ commute. However, we cannot readily compute
their common spectrum because the operator ${\mathcal{S}}_{{\rm
cts}}{\mathcal{U}}$ is not self-adjoint.

We might speculate that the value of the decoherence functional
is maximised for a continuous-time projector that corresponds to
a coarse graining around the classical path. Indeed, if we take
such a generic projection operator $P$, we expect that it should
commute with the operator ${\mathcal{S}}_{cts}{\mathcal{U}}$. In
this context, we noticed earlier that the projection operator
which corresponds to a classical solution $(f,h)$ commutes with
the action operator
\begin{equation}
  [{\mathcal{S}}_{cts}{\mathcal{U}} , P_{(f,h)}] = 0.
\end{equation}
Finally, this argument should be compared with the similar
condition for classical histories:
\begin{equation}
  \{ S_{h} , F_{C} \} (\gamma_{cl}) = 0.
\end{equation}

\section*{Conclusions}
We have examined the example of the simple harmonic oscillator,
in one dimension, within the History Projection Operator
formulation of the consistent-histories scheme. We defined the
action operator as the quantum analogue of the classical
Hamilton action functional and we have proved its existence by
finding a representation on the
${\mathcal{F}}({\mathcal{L}}^{2}({\mathbf{R}}))$ space of the
history algebra. We have shown that the action operator is the
generator of two types of time transformations: translations in
time from one Hilbert space ${\mathcal{H}}_{t}$, labeled by the
time parameter $t$, to another Hilbert space with a different
label $t$, and phase changes in time with respect to the time
parameter $s$ of the standard Heisenberg-time evolution that
acts in each individual Hilbert space ${\mathcal{H}}_{t}$. We
have expressed the action operator in terms of the Liouville and
Hamiltonian operators---which are the generators of the two
types of time transformation---and which correspond to the
kinematics and the dynamics of the theory respectively.

We have constructed continuous-time classical histories defined
on the continuous Cartesian product of copies of the phase space
and demonstrated an analogous expression to the classical
Hamilton's equations.

Finally we have shown that the action operator commutes with the
defining operator of the decoherence functional, thus appearing
in the expression for the dynamics of the theory, as would have
been expected.

One of the major reasons for undertaking this study was to
provide new tools for tackling the recalcitrant problem of
constructing a manifestly covariant quantum field theory in the
consistent histories formalism. Work on this problem is now in
progress with the expectation that the Hamiltonian and Liouville
operators will play a central role in the proof of explicit
Poincar\'e invariance of the theory.

\section*{Acknowledgements}
I would like to thank  Chris Isham for the help and support
during this work. I would also like to thank Charis Anastopoulos
and Nikos Plakitsis. I gratefully acknowledge support from the
NATO Research Fellowships department of the Greek Ministry of
Economy.

\end{document}